\documentclass[12pt]{article}
\usepackage{geometry}                
\usepackage{amsmath}
\usepackage{amsfonts}
\usepackage{amssymb}
\usepackage{graphicx}
\usepackage{amssymb}
\usepackage{epstopdf}
\def\call{{\cal L}}
\def\callk{{\cal L}^{(k)}}

\def\b2hat{ {\hat b}_2 }

\def\ak{A^{(k)} }

\def\kmax{k_{max}}

\begin{document}

\begin{titlepage}
\vfill
\begin{flushright}
ACFI-T16-13\\
\end{flushright}

\vskip 3.0cm
\begin{center}
\baselineskip=16pt
%

{\Large\bf  Extended First Law for Entanglement Entropy in Lovelock Gravity}

\vskip 10.mm

{\bf  David Kastor${}^{a}$, Sourya Ray${}^{b}$, Jennie Traschen${}^{a}$} 

\vskip 0.4cm
${}^a$ Amherst Center for Fundamental Interactions, Department of Physics\\ University of Massachusetts, Amherst, MA 01003\\

\vskip 0.1in ${}^b$ Instituto de Ciencias F\'{\i}sicas y Matem\'{a}ticas\\ Universidad Austral de
Chile, Valdivia, Chile\\
\vskip 0.1 in Email: \texttt{kastor@physics.umass.edu, ray@uach.cl, traschen@physics.umass.edu}
\vspace{6pt}
\end{center}
\vskip 0.2in
\par
\begin{center}
{\bf Abstract}
 \end{center}
\begin{quote}
The first law for the holographic entanglement entropy of spheres in a boundary CFT with a bulk Lovelock dual 
is extended to include variations of the bulk Lovelock coupling constants. Such variations in the bulk correspond to perturbations within a family of boundary CFTs.   The new contribution to the first law is found to be the product of the variation $\delta a$ of the A-type
trace anomaly coefficient for even dimensional CFTs, or more generally its extension $\delta a^*$ to include odd dimensional boundaries,  times the ratio $S/a^*$.  
Since $a^*$ is  a measure of the number of degrees of freedom $N$ per unit volume of the boundary CFT,  this new term has the form $\mu\delta N$, where the chemical potential $\mu$ is given by the entanglement entropy per degree of freedom.

   \vfill
\vskip 2.mm
\end{quote}
\hfill
\end{titlepage}

\section{Introduction}

The AdS/CFT correspondence \cite{Maldacena:1997re} has been most extensively studied for CFTs that have bulk Einstein duals.  However, 
this does not include the most general CFTs of interest.  In four dimensions, for example, the trace anomaly for a general CFT  is given by
\begin{equation}
\langle T_a{}^a\rangle = {c\over 16\pi^2} C_{abcd}C^{abcd} - {a\over 16\pi^2}(R_{abcd}R^{abcd}-4R_{ab}R^{ab}+R^2)
\end{equation}
while the holographic calculation of the trace anomaly with Einstein  gravity  in the bulk \cite{Henningson:1998gx} yields only the special case $a=c$.  Studies including higher curvature interactions in the bulk, which allow for more general boundary CFTs, often focus on Lovelock gravity theories \cite{Lovelock:1971yv}, which are better behaved than generic higher curvature theories having, for example, field equations that depend only on the Riemann tensor and not its derivatives\footnote{See, for example, the studies of causality constraints on the bulk higher curvature  duals of CFTs \cite{Brigante:2007nu,Brigante:2008gz,Buchel:2009tt,Hofman:2009ug,deBoer:2009pn,Camanho:2009vw,Buchel:2009sk,deBoer:2009gx,Camanho:2009hu,Camanho:2013pda}.}.

Interest in CFTs with bulk Lovelock duals extends to holographic computations of entanglement entropy \cite{deBoer:2011wk,Hung:2011xb,Ogawa:2011fw,Myers:2012ed,Bhattacharyya:2013jma,Chen:2013rcq,Dong:2013qoa,Camps:2013zua,Pal:2013fha,Banerjee:2014oaa}.
For theories with bulk Einstein duals, Ryu and Takayanagi \cite{Ryu:2006bv} proposed that the entanglement entropy $S_{E}$ associated with the division of the boundary into complementary regions $A$ and $B$ is given by the Bekenstein-Hawking entropy formula
\begin{equation}\label{entropy}
S_E={A_\Sigma\over 4 G}
\end{equation}
where $A_\Sigma$  is the area of a bulk minimal surface $\Sigma$  that is homologous to the boundary between $A$ and $B$.  For CFTs with bulk Lovelock duals, it has been conjectured \cite{Hung:2011xb} that the entanglement entropy will be similarly given by
\begin{equation}\label{entropyL}
S_E=S_L
\end{equation}
where $S_L$ denotes the formula for horizon entropy in Lovelock gravity found by Jacobson and Myers \cite{Jacobson:1993xs}\footnote{The Wald entropy formula \cite{Wald:1993nt,Iyer:1994ys} for Lovelock gravity differs from the Jacobson-Myers formula by extrinsic curvature terms that vanish for the bifurcation surface of a Killing horizon, but not necessarily for a bulk entangling surface.}
evaluated for a surface $\Sigma$, homologous to the boundary between the regions $A$ and $B$, that minimizes $S_L$.  We will assume that this equality holds below and simply denote the entanglement entropy by $S$ in the following.

Entanglement entropy is not a thermal phenomenon.  However, it has been shown to obey a first law with respect to variations in the quantum state of the CFT  \cite{Blanco:2013joa,Wong:2013gua}.  For a spherical entangling surface on the boundary, this first law follows from the bulk gravitational first law associated with the entangling surface $\Sigma$ \cite{Faulkner:2013ica}.  This works because the bulk surface $\Sigma$, in this case, turns out to be the bifurcation surface of a Killing horizon.   The proof of the first law for stationary black holes \cite{SudarskyWald:1992} then applies in this non-black hole setting as well. 


In reference \cite{Kastor:2014dra}, we used the bulk methods of \cite{Faulkner:2013ica} to prove an extension of the entanglement first law \cite{Blanco:2013joa,Wong:2013gua} for CFTs with a bulk Einstein dual, that gives the variation in entanglement entropy with respect to variation in the number of CFT degrees of freedom\footnote{As in \cite{Faulkner:2013ica}, our construction applies only to spherical entangling surfaces on the boundary, but may well hold more generally.}.  This result relies on the earlier  generalization of the bulk first law 
to include variations in the cosmological constant $\Lambda$ \cite{Kastor:2009wy}.  For static black holes, this latter result has the form
\begin{equation}
\delta M = {\kappa\delta A\over 8\pi G}- {V\delta\Lambda\over 8\pi G}
\end{equation}
where $M$ is the ADM mass, $A$ the horizon area, $\kappa$ the surface gravity.  The quantity $V$ is the `thermodynamic volume' of the black hole, which is conjugate to the cosmological constant $\Lambda$, which can be regarded as a pressure\footnote{The additional term in the first law then has a $V\delta P$ form.  Including this additional part of the black hole phase space has led to a rich phenomenology of phase transitions (see \cite{Kubiznak:2014zwa} for a review of progress in this area).}. 
In the AdS/CFT correspondence, with Einstein gravity in the bulk, the cosmological constant is a measure of the number of CFT degrees of freedom.  For example, for $AdS_5/CFT_4$ \cite{Maldacena:1997re} the number of degrees of freedom per unit volume of the boundary 
$\mathcal{N}=4$ $SU(N)$ gauge theory scales as $N^2$ and  is related to the bulk cosmological constant by
\begin{equation}
N^2 ={\pi\over 2 G_5}\left(-{6\over \Lambda}\right)^{{3\over 2}}
\end{equation}
Similarly in $AdS_3/CFT_2$, the CFT central charge $c$ is a measure of the number of degrees of freedom and is given \cite{Brown:1986nw} in terms of the bulk cosmological constant by 
%
\begin{equation}
c= {3\over 2G_3\sqrt{-\Lambda}}
\end{equation}
Varying the bulk cosmological constant corresponds to varying the number of boundary degrees of freedom, and this forms the basis for the extension of the first law of entanglement entropy in \cite{Kastor:2014dra}.  The quantity conjugate to varying the number of boundary degrees of freedom has a natural interpretation as a chemical potential.

The connection between the bulk gravitational first law and the first law for entanglement entropy, for spherical entangling surfaces, also holds in Lovelock gravity \cite{Faulkner:2013ica}.  An extension of the bulk first law for black holes to include variations in the Lovelock couplings was proved in \cite{Kastor:2010gq}.  In addition to the thermodynamic volume, there are now  thermodynamic potentials conjugate to each of the higher curvature couplings.  The purpose of this paper is to apply this extended bulk first law to derive an extension of the first law for entanglement entropy in this theory, for the case of spherical entangling surfaces.  Variation in the bulk Lovelock coefficients corresponds to variation within a family of boundary CFTs.   We find that at fixed energy, the variation in CFT entanglement entropy with respect to the bulk Lovelock coupling constants assembles into the simple form
  \begin{equation}\label{intros}
  \delta S =  S\, { \delta a^* \over a^* }
 \end{equation}
where the quantity $a^*$ can alternatively be viewed as a function of the Lovelock couplings or in terms of its significance in the corresponding boundary CFT.   For even dimensional CFTs, $a^*$ is the suitably normalized coefficient of the Euler density term in the trace anomaly, expressed in terms of the bulk Lovelock couplings \cite{Liu:2010xc}.  This expression can then be continued to odd dimensions as well.
It has been argued that $a^*$ is proportional to the density of degrees of freedom in the CFT \cite{Myers:2012ed,Liu:2012eea}, and so the effect of
including the variation of the Lovelock couplings is to give a work term in the first law accounting for the change in the number of
degrees of freedom, with
 chemical potential  proportional to $S / a^*$, which can be interpreted as the entanglement entropy per degree of freedom.

 This paper is organized as follows. In Section 2 we present the first law in terms of the entanglement entropy and anomaly coefficient. In Section 3
we give the derivation of the first law in terms of the area and Lovelock coefficients, which is the basis for the 
results of Section 2. In section 4 we offer some concluding remarks.

\section{Extended First Law for Entanglement Entropy}\label{results}

We consider entanglement entropy in CFTs with bulk Lovelock gravity duals. The Lagrangian for Lovelock gravity  in $D$ spacetime dimensions is given by 
\begin{equation}\label{lovelocklagrangian}
{\cal L}={1\over 16 \pi G} \sum_{k= 0}^{\kmax}b_k{\cal L}^{(k)}
\end{equation}
where $\kmax = [(D-1)/2]$ and the $b_k$ are real-valued coupling constants.  The symbol
$\callk$ stands for the contraction of $k$ powers of the Riemann tensor given by
\begin{equation}\label{lovelagran}
\call^{(k)} ={1\over 2^k } \delta ^{a_1 b_1...a_k b_k } _{c_1 d_1 ....c_k d_k }
 R_{a_1 b_1}{}^{c_1 d_1 }\dots  R_{a_k b_k}{}^{c_k d_k }.
\end{equation}
Here the $\delta$-symbol is the totally anti-symmetrized product
normalized so that it takes nonzero values $\pm 1$.  
The term $\call^{(0)}$ gives the cosmological constant term in the Lagrangian, while $\call^{(1)}$ gives the Einstein-Hilbert term and $\call^{(2)}$ the quadratic Gauss-Bonnet term. The upper bound in the sum (\ref{lovelocklagrangian}) comes about because $\call^{(k)}$ vanishes identically for $D<2k$ and turns out to make no contribution to the equations of motion in $D=2k$.  We will fix $b_1 =1$ and note that $b_0 = -2\Lambda$, where $\Lambda$ is the cosmological constant.

Let $\xi$ be a Killing field with a bifurcate Killing horizon, and let  $\Sigma$ be the intersection of the horizon with a constant time slice. Then
the entropy associated with $\Sigma$ is a sum of contributions associated
with each curvature term in the Lovelock Lagrangian, given by  \cite{Jacobson:1993xs} 
\begin{equation}\label{sdef}
S={1\over 4G} \sum b_k A^{(k)} \  , \quad  \ak = k\int_\Sigma da \call ^{(k-1)} [\gamma ]
\end{equation}
where $\gamma_{ab}$ is the induced metric\footnote{We have omitted the  boundary term that appears in the definition of $S$ in reference   
\cite{Jacobson:1993xs} 
since this vanishes when $\Sigma$ is generated by a Killing field. } on $\Sigma$. The $k=1 $ term is just the area, while the $k=0$ term vanishes, corresponding to the fact that the cosmological constant term in the Lagrangian does not 
contribute to  the entropy. The first law in Lovelock gravity including variations of the Lovelock parameters \cite{Kastor:2010gq} is given by 
\begin{equation}\label{lfirst}
\delta E = {\kappa\over 2\pi} \delta S -{1\over 16\pi G}\left( 2 V\delta \Lambda  +  \sum_{k=2} \Psi ^{(k)} \delta b_k \right)
\end{equation}
Here $V$ is the thermodynamic volume, the parameter conjugate to $\Lambda$ \cite{Kastor:2009wy}, the $\Psi ^{(k)}$ are potentials
conjugate to the higher curvature couplings $b_k$ with $k\ge 2$, and $E$ is the ADM charge associated with the Killing field.
One can wonder about the motivation for varying the Lovelock couplings $b_k$.    In fact, we will see that
varying the couplings is the right thing to do in order to compute the chemical potential for a dual CFT, or equivalently,
 the change in the entanglement entropy due to a variation in the `A'-type anomaly coefficient.  
 
We will be working in an asymptotically $AdS$ spacetime, such that the metric in the asymptotic regime is given approximately in Poincare coordinates by  
\begin{equation}\label{metric }
ds^2 \approx  {l^2 \over z^2 }\left( - dt^2 +  dz^2 + dr^2 + r^2 d\Omega^2_{D-3}  \right)
\end{equation}
 where $l$ is the AdS curvature scale. Consider the Killing field used in \cite{Faulkner:2013ica}
\begin{align}\label{Killingvector}
\xi=-\dfrac{2\pi t }{r_0}(z\partial_z+r\partial_r)+\dfrac{\pi}{r_0}(r_0^2-z^2-r^2-t^2)\partial_t
\end{align}
At time $t=0$, the horizon of $\xi$ 
is given by the surface $\Sigma : \ z^2 + r^2 = r_0 ^2$.  The surface $\Sigma$
intersects the AdS boundary  at $z=0$ in a $(D-3)$-sphere $r^2 =r_0 ^2$
whose interior is, in turn, a $(D-2)$-dimensional ball  $B$ on the boundary.  Because $\Sigma$ is a bifurcation surface for the Killing vector $\xi$, the first law (\ref{lfirst}) applies
for perturbations about AdS, and on the other hand,
since $\Sigma$ is a minimal surface one can apply holographic conjecture \cite{Ryu:2006bv} to relate the area of $\Sigma$ to the 
entanglement entropy of the boundary sphere, as follows.

First we review relevant features of the first law results for Einstein gravity \cite{Faulkner:2013ica} 
including the extension to include variation in $\Lambda$ \cite{Kastor:2014dra}.  In this case, the entropy $S$ in the first law  reduces to $A/4G$, where $A$ is the area of $\Sigma$ and one finds that $V_{therm}$ is also proportional to $A$, given by
\begin{equation}\label{vtherm}
V= {2\pi l^2 \over D-1 } A
\end{equation} 
Note that
in these and subsequent formulas $A$ denotes the regularized area, obtained by cutting off the area integral at some small value $z=\epsilon$,
 since the area receives an infinite contribution from the region near the AdS boundary.  This divergence of the area $A$ as $\epsilon\rightarrow 0$ corresponds to the divergence of the entanglement entropy in the boundary CFT as a cutoff is removed. The surface gravity for the Killing vector $\xi$ is found to be $\kappa=2\pi$, and the first law (\ref{lfirst}) with the higher curvature terms set to zero then reduces to
\begin{equation}\label{firsta}
 \delta E =  {\delta A \over 4G} - (D-2) {A\over 4G} {\delta l \over l } 
 \end{equation}
where the cosmological constant $\Lambda$ has been rewritten in terms of the $AdS$ curvature scale by means of 
 \begin{equation}\label{lambdal}
 \Lambda = -{(D-1)(D-2) \over 2 l^2 } 
 \end{equation} 
For $AdS_5$ the dual CFT is  given by ${\cal N} =4 \  SU(N)$ Super-Yang-Mills theory \cite{Maldacena:1997re}, where the $AdS$  curvature scale is related to the number of colors $N$ according to $l^8 \sim N^2$.  The first law (\ref{firsta})  can then be written in terms of variations in $N$ as
\begin{equation}\label{firstn}
 \delta E = \delta S - (D-2)  {S\over N^2 }  \delta (N^2 ) 
 \end{equation}
The number of degrees of freedom of the CFT is proportional to $N^2$, and therefore (\ref{firstn}) determines
the chemical potential for changing the number of degrees of freedom of the boundary CFT to be
\begin{equation}\label{einstchem}
\mu = -(D-2){S\over N^2 }
\end{equation}
Hence, including $\delta \Lambda$ in the first law has allowed us to identify the chemical potential $\mu_{chem}$, which is seen 
to be proportional to the entanglement entropy per degree of freedom. We note that other work has also included a temperature associated
with the variation of $E$ with respect to $S$ \cite{Bhattacharya:2012mi,Guo:2013aca}.   

We find that a similar result holds for boundary CFTs that are dual to Lovelock gravity in the bulk. 
The derivation of this result will be given in section (\ref{derivation}) below.  Here we will focus on the results.
A key feature of the calculation given below is that with the constant curvature form of the Riemann tensor for $AdS$, each of the terms in first law (\ref{lfirst}) works out to be proportional to the corresponding term in the Einstein case.   Consequently, a sum over the Lovelock coupling constants $b_k$ factors out of the entire equation, giving a simple result in terms
of the horizon area.  The extended first law including variations in the Lovelock couplings will then take a simple form if we rescale the variation of the energy according to
  \begin{equation}\label{etilde}
 \delta \tilde{ E} = {(D-1)(D-2) \over s_{(1)} }\delta {E} 
  \end{equation}
where the quantity $s_{(1)}$ is given by the sum over the Lovelock couplings
 \begin{equation}\label{sone}
s_{(1)}=- l^2 (D-1)!  \sum_{k=0}{ (-1)^k \over l^{2k} } {k\  b_k \over (D-2k-1)!  }
  \end{equation}
Written in terms of rescaled quantity $\delta\tilde E$, the extended first law in Lovelock gravity for perturbations about the minimal surface $\Sigma$ that intersects the $AdS$ boundary in a sphere is then given by
\begin{equation}\label{firstal}
  \delta \tilde{ E}=  {\delta A \over 4G} - (D-2) {A\over 4G} {\delta l \over l } 
 \end{equation}
where the $AdS$ curvature scale $l$ is now related to the Lovelock couplings by
 \begin{equation}\label{szero}
  \sum_{k=0}{ (-1)^k \over l^{2k} } {b_k \over (D-2k-1)!  } =0
 \end{equation}
 Note that this result has the same form as the first law (\ref{firsta}) with Einstein gravity in the bulk.

In Einstein gravity, the first law written in terms of the horizon area (\ref{firsta}) translates directly into a statement (\ref{firstn}) about the entropy and its variation.
However, in Lovelock gravity such a restatement requires additional steps.
The different horizon integrals contributing the entropy (\ref{sdef}) for the surface $\Sigma$ 
all work out to be proportional to its area, with the 
$A^{(k)}$  given by
\begin{equation}\label{ak}
\ak = { A\over 4G}\, k\,  e_k  \  , \qquad    e_k = \left( -1 \over l^2 \right)^{k-1} {(D-2)! \over (D-2k)! } 
\end{equation}
Substituting this in (\ref{sdef}) we find that the entropy associated with the minimal surface $\Sigma$ is given by
\begin{equation}\label{loves}
S=  \left(\sum_{k} k e_k b_k  \right) {A\over 4G} 
\end{equation}

The entropy can be  rewritten in terms of the `$A$'-type anomaly
coefficient $a^* $. In the four-dimensional CFT ${\cal N} =4 \  SU(N)$ super-Yang-Mills
  dual to Einstein gravity without higher curvature terms the central charges $a$ and $c$ are equal. However with higher deivative
  terms in the gravitational Lagrangians this is no longer the case. 
Both holographic and direct CFT calculations of the entanglement entropy have found that the several anomaly coeffients of the CFT
can be distinguished by studying entangling  boundaries with different geometries \cite{Solodukhin:2008dh,Myers:2010tj,Hung:2011xb,deBoer:2011wk}.
For example, in Lovelock gravity it was found that the entanglement entropy of a cylinder is proportional to the `$c$' coefficient, 
while that of a sphere is proportional to $a^*$, where
\begin{equation}\label{astar}
a^*   = {l^{D-2} \over 4G}\left(  \sum_k  k e_k b_k\right)
\end{equation}
 The `star' indicates that the coefficient has been extended to include CFTs of odd dimensions\footnote{Our normalization differs from that in \cite{Hung:2011xb} 
  by the factor $  4\pi ^{(D-1)/2} /[ \Gamma ( {D-1\over 2 }) ]$}. Hence the entropy  \cite{Hung:2011xb,deBoer:2011wk} can be written as
 \begin{equation}\label{lovestwo}
S=  {a^* A \over l^{D-2}}  
\end{equation} 
and its variation is given by
\begin{equation}\label{varys}
\delta S=  {a^* A\over l^{D-2} }  \left(   {\delta a^*\over a^*} -(D-2) {\delta l \over l }   \right) + { a^* \over l^{D-2} } \delta A 
\end{equation}
The prefactor of the first term above is just the entropy, and  the last term can be rewritten  in terms of the entropy and the change in energy using the first law (\ref{firstal}).
 One of the terms in the first
law cancels the $\delta l$ term, and so all the variations of the Lovelock couplings combine to form $\delta a^*$,
giving the result
 \begin{equation}\label{firstsl}
\delta S=    S{\delta a^* \over a^* }   + { 4a^* G \over l^{D-2} }   \delta \tilde{E}  
\end{equation} 
%
The anomaly coefficient $a^*$ has the interpretation of the number of degrees of freedom per cell in the regulated field theory \cite{Myers:2010tj,Myers:2010xs}
and so the last term is proportional to the change in the number of degrees of freedom in the CFT, as was the case
for Einstein-$\Lambda$ gravity (compare to equation (\ref{firstn})). Hence
the variation of $S$ with respect to $a^*$ at fixed energy, for a spherical boundary, can be thought of as a generalized chemical potential with value
\begin{equation}\label{chem}
\mu = \left. {\partial S \over \partial a^* } \right |_E = {S \over a^*}
\end{equation}
for a CFT dual to a Lovelock theory.  The chemical potential $\mu$ is simply the entanglement entropy per degree of freedom.

\subsection{Explicit formula and an example in $D=5$}

The relation (\ref{firstsl}) which gives $\delta S$ at fixed energy in terms of $\delta a^*$ is a nice and compact expression.   However, it is also useful to have the equivalent
expression in terms of the variations of the Lovelock coefficients and the AdS radius.  This is given by
\begin{equation}\label{deltasb}
\left. \delta S \right |_E
=\dfrac{A}{4G}\sum\limits_{k=0}^{\kmax}\left(\dfrac{-1}{l^2}\right)^{k-1}\dfrac{k(D-2)!}{(D-2k)!}\left[\delta b_k+(D-2k)b_k\dfrac{\delta l}{l}\right]\\
\end{equation}
As an illustration and a check of our work, in this section we start with the entropy in terms of the Lovelock couplings and translate to the 
conformal field theory coefficients  $a$ and $c$, calculated by other techniques. In $D=5$ the only nonzero  coupling constants are  $b_0$, $b_1\equiv 1$, and $b_2$, and equation (\ref{deltasb}) reduces to
\begin{align}\label{deltaS}
\left .\delta S \right |_E =-\dfrac{3A}{4G}\left[\dfrac{4\delta b_2}{l^2}+\left(\dfrac{4b_2}{l^2}-1\right)\dfrac{\delta l}{l}\right]
\end{align}
From reference \cite{deBoer:2011wk}, the coupling constants $b_0$ and $b_2$ are related to the $4D$ CFT trace anomaly coefficients $a$ and $c$ according to
\begin{equation}
b_0b_2 =\dfrac{3(a-5c)(a-c)}{2(a-3c)^2}
\end{equation}
while from (\ref{szero}) the AdS radius $l$ is determined by the equation
\begin{align*}
l^2b_0-12+\dfrac{24}{l^2}b_2=0
\end{align*}
These can be combined to obtain expressions for the couplings in terms of $a$, $c$ and $l$
\begin{align}
b_2=\dfrac{l^2(a-c)}{4(a-3c)} \qquad \text{and} \qquad b_0=\dfrac{6(a-5c)}{l^2(a-3c)}.
\end{align}
Additionally the AdS radius is given \cite{deBoer:2011wk} in terms of the anomaly coefficients and Newton's constant by $l^3=G(3c-a)/90\pi$.
Hence the variations of the Gauss-Bonnet coupling $b_2$ and the AdS radius $l$ can be expressed in terms of the variations of the anomaly coefficeints by
\begin{align*}
\dfrac{\delta b_2}{l^2}=\dfrac{(a-c)}{2(a-3c)}\dfrac{\delta l}{l}+\dfrac{a\delta c-c\delta a}{2(a-3c)^2}\ , \qquad \dfrac{\delta l}{l}=\dfrac{3\delta c-\delta a}{3(3c-a)}
\end{align*}
%
Plugging these into the expression (\ref{deltaS}) for the variation of the entanglement entropy gives
\begin{align}
\left .\delta S \right |_E =-\dfrac{A}{2G}\dfrac{\delta a}{(a-3c)}
\end{align}
Note that the terms proportional to $\delta c$ have cancelled. The unperturbed entropy $S$ is determined by (\ref{loves}) to be
\begin{align}\label{entnglentropy}
S&=\dfrac{A}{4G}\left(1-\dfrac{12b_2}{l^2}\right)\\ &=-\dfrac{A}{2G}\dfrac{a}{(a-3c)}
\end{align}
Hence the variation of the entanglement entropy is found to be $\left .\delta S \right |_E =S\dfrac{\delta a}{a}$,
which is in agreement with the general result (\ref{firstsl}) above, since in this case $a^*=a$.

\section{Details of the Derivation}\label{derivation}
In this section we give the details of the derivation of the extended first law (\ref{firstal}) including variations in the Lovelock couplings for the change in the area of the minimal surface $\Sigma$ that intersects the $AdS$ boundary in a sphere.
The extended first law is valid for small perturbations around $AdS$, as well as in the far field of an $AdS$ black hole.
The derivation makes use of the Hamiltonian perturbation theory formalism.
Here only give needed details for the derivation at hand, while a more complete treatment can be found in \cite{Kastor:2010gq}.

In the Hamiltonian framework, we start by decomposing the metric as
\begin{equation}\label{split}
g_{ab} = - n_a n_b +s_{ab} \  , \quad n_a n^a =-1 \  , \quad n_a s^a{}_b =0
\end{equation}
In the asymptotically $AdS$ region the timelike normal is simply $n_a = -(l/z)\nabla _a t$.
 The Killing field used in the first law construction, given explicitly in (\ref {Killingvector}), is 
 decomposed as
 \begin{equation}\label{kvsplit}
 \xi ^a = F n^a + \beta ^a
 \end{equation}
 We will take the background metric to be $AdS$, and denote
 the perturbation to the spatial metric as $ h_{ab}$,
 \begin{equation}\label{hdef}
 s_{ab} = s_{ab}^{AdS} + h_{ab}
 \end{equation}
 There is also a perturbation to the gravitational momentum, but it doesn't enter into this calculation.
 
The extended first law was derived
previously for black holes \cite{Kastor:2010gq}, with the result given above in (\ref{lfirst}). For the geometry of interest here,
it is convenient to backtrack to a more `primitive' version of the first law.  This amounts to 
the following integral identity, which holds for solutions about a background solution to the Lovelock equations of motion
\begin{align}\label{bounding}
\int_{\partial V}da_c\sum_k(b_kB^{(k)c}+\delta b_k\beta^{(k)cd}n_d)=0.
\end{align}
Here the boundary vectors $B^{(k) a}$, which depend on the metric perturbation, are given by
\begin{equation}\label{bk}
B^{(k) a} ={k\over 2^{k-1} } \delta ^{a b a_1 b_1...a_{k-1} b_{k-1} } _{cd c_1 d_1 ....c_{k-1} d_{k-1} }
 R_{a_1 b_1}{}^{c_1 d_1 }\dots  R_{a_{k-1} b_{k-1} }{}^{c_{k-1} d_{k-1} } \left( F\nabla^c h^d_b  -h^d_b \nabla^c F \right)
\end{equation}
and the Killing-Lovelock potentials $\beta^{(k)ab}$  \cite{Kastor:2008xb} corresponding to the Killing vector $\xi^a$ are solutions to
\begin{align*}
-\dfrac{1}{2}\nabla_b\beta^{(k)ba}=\mathcal{G}^{(k)a}_{\ \ \ b}\xi^b
\end{align*}
Here $\mathcal{G}^{(k)a}_{\ \ \ b}$ is the $k$th order Lovelock tensor,
\begin{equation}\label{lgk}
\mathcal{G}^{(k)a}_{\ \ \ b}= - {1\over 2^{k+1} } \delta ^{a a_1 b_1...a_k b_k } _{b c_1 d_1 ....c_k d_k }
 R_{a_1 b_1}{}^{c_1 d_1 }\dots  R_{a_k b_k}{}^{c_k d_k }.
\end{equation}
We apply the identity (\ref{bounding}) with the Killing vector (\ref{Killingvector}) to the boundary composed of 
 the spherical ball $B$ of radius $r_0$  at spatial infinity plus the bulk minimal surface $\Sigma$, which is the bifurcate Killing horizon of $\xi$.
 Since the background is $AdS$ and the Riemann tensor has the simple constant curvature form, it turns out that the various lengthy expressions indexed by $k$ differ only in the multiplicative  pre-factors.  Explicitly,
using $R^{ab}_{cd}=-(1/l^2)\delta^{ab}_{cd}$, one finds the Lovelock tensors (\ref{lgk}) are given by
\begin{align*}
\mathcal{G}^{(k)a}_{\ \ \ b}=-\dfrac{1}{2}\left(\dfrac{-1}{l^2}\right)^k\dfrac{(D-1)!}{(D-2k-1)!}\delta^a_b.
\end{align*}
and that the Killing-Lovelock potentials can be written in terms of the  $k=0$ Killing potential as
\begin{equation}\label{betak}
\beta^{(k)ab}=\left(\dfrac{-1}{l^2}\right)^k\dfrac{(D-1)!}{(D-2k-1)!} \beta^{(0)ab}
\end{equation}
Further, the Killing potential $\beta^{(0)ab}$ can be obtained simply by combining the Ricci identity $\nabla_a\nabla^b\xi^b=-R^b_a\xi^a$ for the Killing vector along with
 the Ricci tensor $R_{ab}=-((D-1)/l^2)g_{ab}$ of the AdS background which gives
\begin{align*}
\beta^{(0)ab}=\dfrac{l^2}{D-1}\nabla^a\xi^b.
\end{align*}
For the Killing vector (\ref{Killingvector}), this gives
\begin{align}\label{Killingpots}
\beta^{(k)}&=\dfrac{1}{2}\beta^{(k)ab}\partial_a\wedge \partial_b\\
&=\left(\dfrac{-1}{l^2}\right)^{k}\dfrac{\pi z(D-2)!}{r_0(D-2k-1)!}\Big\{(r_0^2+z^2-t^2-r^2)\partial_t\wedge\partial_z
+2tx^k\partial_z\wedge\partial_k+2zx^k\partial_t\wedge\partial_k\Big\} 
\end{align}
%
Similarly, the boundary terms $B^{(k)a}$ corresponding to each order $k$ can be expressed in terms of the boundary term corresponding to the Einstein-Hilbert term (i.e., $k=1$) as
\begin{align}\label{numfactor}
B^{(k)a}=k\left(\dfrac{-1}{l^2}\right)^{k-1}\dfrac{(D-3)!}{(D-2k-1)!}B^{(1)a}
\end{align}
\begin{equation}\label{bone}
B^{(1)a} = F(D^a h -D_b h^{ab} ) -hD^a F +h^{ab}D_b F  
\end{equation}
The weighted sum of boundary vectors can be expressed more compactly as
\begin{align}\label{sumbdnryvecs}
\sum_{k}b_kB^{(k)a}=\dfrac{s_{(1)}}{(D-1)(D-2)}B^{(1)a}
\end{align}
where the sum $s_{(1)}$ was defined in (\ref{sone}).

\subsection{Boundary at infinity} We are now ready to 
evaluate the integral (\ref{bounding}) on the boundary at spatial infinity. We will find that at infinity
the terms arising due to variations in the coupling constants $b_k$ are separately divergent but sum to zero,
leaving only the ADM energy corresponding to the Killing field $\xi^a$. This cancellation works essentially in the same
way as in the calculations in \cite{Kastor:2014dra} and \cite{Kastor:2010gq}.
First analyze the boundary vector $B^{(1)a}$ in equation (\ref{bone}), which depends on $h_{ab}$. The metric perturbation can be divided into a contribution
with $l$ held fixed, and a contribution from a change in $l$. The second portion is simply  $h_{ab}=(2\delta l/l)s_{ab}$. The normal component $F$ of the Killing field on the boundary is $F=(\pi l/r_0z)(r_0^2-r^2)$. 
 \begin{align*}
da_z B^{(1)z}= \left( {l^2 \over z^2 }  r dr d\Omega \right) \left[ B^{(1)z} \big\rvert_{\delta l=0}+\dfrac{2(D-2)\pi \delta l}{r_0l^2}(r_0^2 -r^2) \right]
\end{align*} 
We now integrate the sum of boundary vectors in (\ref{sumbdnryvecs}) on the boundary at infinity using the above expression. 
The integral of the term corresponding to the perturbation with the $AdS$ length $l$ held fixed gives the variation in the 
ADM charge associated with the Killing field $\xi^a$ \cite{Kastor:2011qp}, and hence we obtain
\begin{align}\label{intboundvec}
\int_B da_c\sum_{k}b_kB^{(k)c}=-16\pi G \delta E_{\xi}+ 
 \delta l \dfrac{2\pi  s_{(1)}}{(D-1)}  {K\over z^2 }
\end{align}
where
\begin{equation}\label{kdef}
K = {1\over r_0  } \int_B { r dr d\Omega \over z^2 }   (r_0^2 -r^2 ) = \Omega_{D-3} {r_0^3 \over 4}
\end{equation}
 The last term in (\ref{intboundvec}) diverges as $z\rightarrow 0$, which is to be expected from the way
$l$ enters the metric. However, this will be cancelled by the contribution from the Lovelock-Killing potentials, which 
we evaluate next.  The relevant contribution is from the components $\beta^{(k)tz}$, which from (\ref{Killingpots}) gives
\begin{align*}
\sum_{k=0}\delta b_k\beta^{(k)zt}n_t=\left(\sum_{k=0}\left(\dfrac{-1}{l^2}\right)^{k}\dfrac{(D-2)!\delta b_k}{(D-2k-1)!}\right)\dfrac{\pi l}{r_0}(r_0^2 -r^2 )
\end{align*}
The sum inside the parenthesis on the right hand side involving $\delta b_k$'s can be expressed in terms of $\delta l$ by taking
the variation of equation (\ref{szero}),
\begin{align}\label{deltab-deltal}
\sum_{k=0}\left(\dfrac{-1}{l^2}\right)^{k}\dfrac{(D-1)!\delta b_k}{(D-2k-1)!}&=-s_{(1)}\dfrac{2\delta l}{l^3}.
\end{align}
Finally, integrating over the boundary at infinity, we get
\begin{align}
\int_B da_c\sum_{k=0}\delta b_k\beta^{(k)cd}n_d=-   \delta l \dfrac{2\pi  s_{(1)}}{(D-1)r_0}   {K \over z^2}
\end{align}
which precisely cancels the diverging contribution in (\ref{intboundvec}).

\subsection{Boundary in the interior} 

Let us now evaluate the integral in (\ref{bounding}) on the bulk minimal surface $\Sigma$ in the interior. 
Since $\Sigma$  is a bifurcate Killing horizon, the integral of the boundary vector $B^{(1)a}$ over this surface is $-2\kappa \delta A$, where $A$ is the area of the minimal surface and $\kappa=2\pi$ for the Killing vector $\xi^a$. Using (\ref{sumbdnryvecs}), we then have
\begin{align}\label{intbndryvec}
\int_{\Sigma}da_c\sum_{k}b_kB^{(k)c}=-\dfrac{4\pi s_{(1)}\delta A}{(D-1)(D-2)}.
\end{align}
There remains the integral of the Killing-Potential terms on $\Sigma$.
Since  the Killing potentials of different orders differ only in the multiplicative factors as displayed in equation (\ref{betak}),
each of the integrals  is proportional to the thermodynamic volume defined by 
\begin{align*}
V_{ther}  = - \int_{\Sigma}da_c\beta^{(0)cb}n_b
\end{align*}
which, noting that the unit outward normal to $\Sigma$ is $ m=-l (zdz+\vec{x}\cdot d\vec{x}) /(zr_0 )$, has the value
\begin{align}
V_{ther} =\dfrac{2\pi l^2}{D-1}A.
\end{align}

Using (\ref{Killingpots}), we then obtain the sum of all contributions of the Killing potentials on the bulk minimal surface in terms of the area
\begin{align}\label{intKillingpot}
\int_{\Sigma}da_c\sum_{k}\delta b_k\beta^{(k)cd}n_d
= \dfrac{4\pi s_{(1)}A}{D-1}\dfrac{\delta l}{l}.
\end{align}
In the last line we have expressed the sum involving the $\delta b_k$ in terms of $\delta l$ using (\ref{deltab-deltal}). Adding
the contributions on the boundary $\Sigma$, equations (\ref{intbndryvec}) and (\ref{intKillingpot}), gives
\begin{align}
\int_{\Sigma}da_c\sum_{k}\left(b_kB^{(k)c}+\delta b_k\beta^{(k)cd}n_d\right)=-\dfrac{4\pi s_{(1)}}{(D-1)(D-2)}\left[\delta A-(D-2)A\dfrac{\delta l}{l}\right].
\end{align}
Combining the results from the previous subsections we have the following extension of the first law for an entangling surface
that intersects the AdS boundary in a sphere,
\begin{align*}
\delta E_{\xi}=\dfrac{s_{(1)}}{4G(D-1)(D-2)}\left[\delta A -(D-2)A\dfrac{\delta l}{l}\right]
\end{align*}
which completes the derivation for (\ref{firstal}) which forms the input in Section (\ref{results}) leading to our main result (\ref{firstsl}).

\section{Conclusion}
To summarize, by making use of standard Hamiltonian perturbation theory methods we have derived an extended form of the first law for entanglement entropy, for spherical entangling surfaces, in CFTs with a bulk Lovelock dual.  This extension  gives the variation of the holographic entanglement entropy as the bulk Lovelock coupling constants are varied, corresponding to variation within a family of boundary CFTs.
We have shown that variations of the bulk Lovelock couplings impact the entanglement entropy through their contributions to the variation of the A-type trace anomaly coefficient $a$ of the boundary CFT, or its generalization $a^*$ for odd dimensional boundaries.
At constant energy, we find that the logarithmic  change in $S$ is equal to the logarithmic change in $a^*$.   Given that $a^*$ is a measure of the number of degrees of freedom of the CFT, we can regard the quantity $S/a^*$ as a chemical potential for increasing the number of degrees of freedom within a family of boundary CFTs.

One natural question is whether the variation in holographic entanglement entropy with respect to variations in the bulk Lovelock couplings is linked to the variation in the trace anomaly coefficients flow for more generally shaped regions, such as strips.  A second, more speculative question is whether there might be connection between a bulk second law associated with entangling surfaces and renormalization group flow between UV and IR CFTs.
A great deal of work has been done on the issue of a higher dimensional
 version of the $c$-theorem \cite{Zamolodchikov,Freedman:1999gp,Solodukhin:2006ic,Myers:2010tj,Myers:2010xs,Liu:2010xc,Liu:2011iia, 
 Casini:2012ei,Komargodski:2011xv,Komargodski:2011vj,Solodukhin:2013yha,Bhattacharjee:2015qaa}. 
References  \cite{Myers:2010tj,Myers:2010xs} showed
 that with an energy condition, $ (a^* )_{UV} \geq (a^* ) _{IR}$, and so based on equation (\ref{intros}) 
 one can speculate that an entropy increase property is connected to the anomaly flow. In this context, interesting work on entropy increase has appeared  in \cite{Bhattacharjee:2015qaa,Bhattacharjee:2015yaa,He:2014lfa}.

\subsection*{Acknowledgements}  The work of S.R. is supported by FONDECYT grant 1150907. S.R. would also like to thank the Amherst Center for Fundamental Interactions at UMass, Amherst for their hospitality, and Adrian and Valerie Parsegian for the charming accommodations at the Casa F\'isica in Amherst during his visit while part of this work was carried out.

\end{document}